\documentclass[twoside]{pnastwo}

\usepackage{amsmath,amssymb,amsfonts,slashed}
\usepackage{pnastwof}
\usepackage[all]{xypic}

\newcommand{\loopx}{\mathcal{L}}
\DeclareMathOperator{\Ind}{Ind}
\DeclareMathOperator{\U}{U}
\DeclareMathOperator{\chern}{ch}
\DeclareMathOperator{\Tr}{Tr}
\DeclareMathOperator{\IND}{IND} \let\IND=\Ind
\DeclareMathOperator{\ind}{ind}
\DeclareMathOperator{\Aroof}{\hat{A}}
\DeclareMathOperator{\ch}{ch}
\DeclareMathOperator{\echern}{sch}
\DeclareMathOperator{\chernchar}{ch}
\DeclareMathOperator{\pontx}{p}
\newcommand{\Dirac}{\slashed{D}}
\newcommand{\bbZ}{\mathbb{Z}}
\newcommand{\nred}{\hat{n}}
\newcommand{\pont}{\pontx_{1}}
\newcommand{\pred}{\hat{p}}
\newcommand{\mred}{\widehat{m}}
\newcommand{\sgenus}{\hat{s}}
\newcommand{\aroof}{\hat{a}}
\newcommand{\susy}{\overline{G}_{0}}
\def\bra#1{\left\langle #1\right\rvert}
\def\ket#1{\left\lvert #1\right\rangle}
\def\braket#1#2{\left\langle#1\vphantom{#2}\right\rvert\!\left.#2\vphantom{#1}\right\rangle}
\begin{document}
\title{The analytic index for a family of Dirac-Ramond operators}
\author{%
Orlando Alvarez%
\affil{1}{Dept. of Physics,  University of Miami}%
\and%
Paul Windey%
\affil{2}{Lpthe Cnrs UMR 7589, Universit\'{e} Pierre et Marie Curie}%
}
\contributor{Submitted to Proceedings of the 
National Academy of Sciences of the United States of America}
\maketitle
\setcounter{page}{1}
\begin{article}
\begin{abstract}
We derive a cohomological formula for the analytic index of the
Dirac-Ramond operator and we exhibit its modular properties.
\end{abstract}

\dropcap{T}he Atiyah-Singer index theorem is important in many
different areas of mathematics and is also at the heart of many
problems in physics.  In quantum field theory it comes into play
through the Dirac operator.  Both the ordinary index
\cite{Atiyah-Singer-I:1968}, and the families index
\cite{Atiyah-Singer-IV:1971} of the Dirac operator reveal features of
quantum field theories.  The Dirac-Ramond operator is the extension to
superstring theory of the ordinary Dirac operator in field theory; it
is the Dirac operator on loop space and its ordinary
index~\cite{alvarez:santa-barbara,Alvarez:1987p1608,%
  Witten:1987p2039,Alvarez:1987p2438} is given by the string genus, while the
elliptic genus of Ochanine and Landweber and
Stong~\cite{Ochanine:1987p1924,Landweber:1988p1980} corresponds to the
index of one of its twisted version.  Both the string genus and the elliptic
genus have been extensively studied.  They have given rise to an
extension of the rigidity theorems of the elliptic genus to the case of
families~\cite{Liu:2000p3725} and to different elliptic cohomology
theories, most notably to the theory of topological modular forms
(tmf) \cite{Ando:2001p3432}.  However, the analytic families index of the
Dirac-Ramond operator has not yet been formulated and in this paper we
derive for it a cohomological formula, with remarkable modular
properties, using methods from field theory and string theory.

Consider a family of Dirac operators parametrized by a space $X$.  The
zero modes of the Dirac operator define a virtual vector bundle $\Ind$
over $X$, called the index bundle.  One of the outcomes of the
families index theorem is a cohomological expression for the Chern
character of this bundle.  The first two terms in the expansion of
this Chern character in characteristic classes are the dimension of
the vector bundle (i.e. the ordinary index of the Dirac operator) and
its first Chern class respectively.  In the Dirac-Ramond case, the
analytic index is not an integer but a modular (or nearly modular)
function whose Fourier coefficients are integers. 

For simplicity, the gist of the argument and the exposition of the
methods will be presented in the case of the Dirac operator.
\emph{Mutatis mutandis} they generalize straightforwardly to the
Dirac-Ramond operator.  It obtains, \emph{en passant}, a novel
presentation of the cohomological Dirac families index theorem that
displays its intimate relation to Berry's
phase~\cite{Berry:1984p2222}.  When needed we will revert to a full
discussion of the Dirac-Ramond case, which is our main concern.

\subsection*{Background.}
\label{sec:background}

The treatment of the Atiyah-Singer index theorem with quantum field
theory techniques has become well known
\cite{AlvarezGaume:1983p1985,Friedan:1984p1942,Witten:AS}.  It rests
on a few general principles.  Consider a supersymmetric quantum
system. The spectrum of its hamiltonian consists of bosonic and
fermionic states. The bosonic and fermionic eigenstates of non zero
energy are paired with each other by supersymmetry.  The generator of
supersymmetry is a Dirac-like operator which commutes with the
hamiltonian and anticommutes with fermion parity. The analytical index
of this Dirac operator, \emph{i.e.} the difference between the number
of its bosonic and fermionic zeros modes, is then easily computed as
the supertrace of the quantum evolution operator. The usual
correspondence between the hamiltonian and the lagrangian formulation
of quantum mechanics gives an equality between this trace and a
supersymmetric path integral that is evaluated by the stationary phase
approximation which is exact and shows that the path integral
localizes. This gives the expression for the topological index.  In
this language, twisting the Dirac operator by some vector bundle
simply amounts to coupling the quantum system to additional degrees of
freedom. The canonical quantization of these degrees of freedom
produces the required vector bundle. In this case, the A-roof genus is
simply replaced by the product of the A-roof genus with the Chern
character of the vector bundle.  Similarly, the equivariant cases
corresponds to having additional symmetries in the supersymmetric
quantum system. Any special case of the Atiyah-Singer
theorem can be treated in this manner.

In their paper on the families index theorem, Atiyah and Singer
\cite{Atiyah-Singer-IV:1971} consider a family of elliptic operators
parametrized by a compact topological space. Consider a smooth family
of metrics on a riemannian spin manifold $Y$ parametrized by a space
$X$.  The manifold $Y$ with metric parametrized by $x\in X$ will be
denoted by $Y_{x}$.  The idea is to put $X$ and $Y$ together into a
fiber bundle $Z \to X$ where at each point $x\in X$ the fiber over $x$
is a manifold isomorphic to $Y$ with metric $g_{Y}(x,\cdot)$.  It is
well known that the index of $\slashed{D}^{Y_{x}}$ is independent of
the metric $g_{Y}(x,\cdot)$.  However the Dirac operator
$\slashed{D}^{Y_{x}}$ and its zero modes change with the metric.
Following Grothendieck, Atiyah and Singer asked how the space of zero
modes changes as $x$ varies over $X$.  At each $x$, the space of zero
modes is a finite dimensional vector space.  Roughly speaking, this
means that the space of zero modes of $\slashed{D}^{Y_{x}}$ is a
finite dimensional vector bundle over $X$.  This is not quite correct
because the dimensionality of the vector space will jump if the number
of solutions to the equation $\slashed{D}^{Y_{x}}\psi = 0$ changes
with $x$.  Only the index of the operator is protected from these
jumps.  If $\mathcal{Z}^{Y_{x}}_{\pm}{}$ are the vector spaces of
respectively positive and negative chirality solutions to the Dirac
equation $\slashed{D}^{Y_{x}}\psi = 0$ with metric $g_{Y}(x,\cdot)$
then $\ind(\slashed{D}^{Y_{x}}) = \dim \mathcal{Z}^{Y_{x}}_{+}{} -\dim
\mathcal{Z}^{Y_{x}}_{-}{}$ is independent of $x$.  Atiyah and Singer
show that the virtual vector spaces $\mathcal{Z}^{Y_{x}}_{+}{} \ominus
\mathcal{Z}^{Y_{x}}_{-}{}$ can be put together over $X$ to make a
virtual vector bundle $\IND(\slashed{D}^{Y})$ over $X$, \emph{i.e.} an
element of $K-$theory, called the index bundle.  Because it is well
suited to the methods of quantum field theory and the study of the
string genus, we consider here a more restrictive family given by a
riemannian submersion.  A riemannian submersion is a family of metrics
with a special relationship between the geometries of $Z$ and $X$.
Pick a point $z \in Z$ that projects to $x\in X$.  At $z$ there is an
orthogonal decomposition of the tangent space $T_{z}Z = H_{z} \oplus
V_{z}$.  Here $V_{z} \subset T_{z}Z$ is the ``vertical subspace''
consisting of vector that are parallel to the fiber.  The ``horizontal
subspace'' $H_{z}$ is the orthogonal complement of $V_{z}$.  A vector
$v \in T_{x}X$ has a unique horizontal lift to a vector $\tilde{v} \in
H_{z}$.  The condition for a riemannian submersion is that $\lVert v
\rVert_{X} = \lVert \tilde{v} \rVert_{Z}$ for every $z
\in\pi^{-1}(x)$.  Note that the restriction of the metric on $Z$ to a
vertical subspaces gives a metric on $Z_{x} \approx Y_{x}$.  If
$x^{i}$ are local coordinates on the base $X$ and if $y^{a}$ are local
coordinates on the fiber $Y$ then $(x,y)$ are local coordinates on
$Z$.  The fibers are the submanifolds with $x$ fixed.  The metric of a
submersion is locally of the ``Kaluza-Klein'' form
\begin{multline}
	ds^{2}_{Z} = g_{ij}(x) dx^{i}\, dx^{i} +\\
	g_{ab}(x,y) \left(dy^{a} + 
	C^{a}{}_{i}(x,y)dx^{i}\right) \left(dy^{b} + 
	C^{b}{}_{j}(x,y)dx^{j} \right)\,.
    \label{eq:metric-Z}
\end{multline}
The metric above leads to a supersymmetric
lagrangian on $Z$ with the following schematic form $L_{Z}(x,y) =
L_{X}(x) + L_{Y}(x,y)$ where $L_{X}$ is the pullback of the
supersymmetric lagrangian on the base and $L_{Y}$ is roughly the
lagrangian on $Y$ depending parametrically on $x$.

\subsection*{Outline of the argument.} Since the original paper of 
Atiyah and Singer there have been a number of different proofs of the families
index theorem. In what follows we present yet another derivation which
has the remarkable feature that its generalization to loop space
obtains for the first time a families index theorem for the
Dirac-Ramond operator.  The multiplicative property of the
index~\cite{Atiyah-Singer-IV:1971} is central to our argument.  It
states that for a submersion $Z \to X$, $\ind \slashed{D}_{Z}$ is the
index of the Dirac operator on $X$ twisted by the index bundle of
$\slashed{D}_{Y}$.  In our setup this multiplicative property is a
reflection of Fubini's integration theorem.  It is important to keep
in mind that our approach is geometrical rather than topological.
This leads to de Rham cohomology represented by differential forms and
thus, in our final formula, we lose all torsion phenomena that might
be of interest.  From now on the expression ``index theorem'' (and
variants thereof) is shorthand for ``cohomological form of the index
theorem'' (and respective variants).

The following diagram displays clearly the architecture of our
arguments:
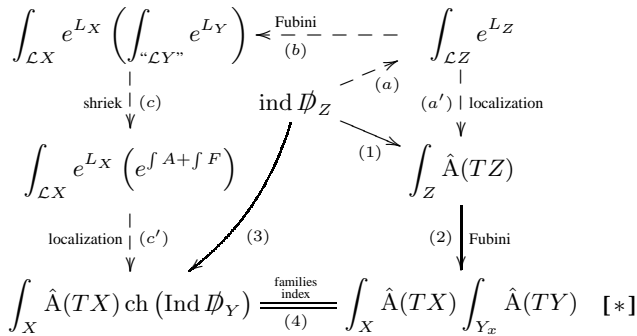
\begin{equation}
    \tag{$*$}
    \label{eq:diag}
\xy
(0,0)*+{\displaystyle \int_{X} \Aroof(TX) \ch\left(\Ind
	\Dirac_{Y}\right) }="C";
(0,18)*++{\displaystyle \int_{\loopx X} e^{L_{X}} \left(e^{\int A + \int
	F}\right)}="L";
(44,18)*+{ \displaystyle 	\int_{Z} \Aroof(TZ)}="R";
(44,36)*+{\quad \displaystyle \int_{\loopx Z} e^{L_{Z}}}="B";
(22,27)*+{\displaystyle\ind \Dirac_{Z}}="O";
(44,0)*+{\displaystyle \int_{X} \Aroof(TX) \int_{Y_{x}}\Aroof(TY)}="D";
(0,36)*+{\displaystyle \int_{\loopx
	X} e^{L_{X}} \left( \int_{\text{``$\loopx Y$''}} e^{L_{Y}}
	\right)}="A";
{\ar@{-->}^{\text{localization}}_{(a')} "B";"R" };
{\ar^{\text{Fubini}}_{(2)} "R";"D" };
{\ar_{(1)} "O";"R" };
{\ar@{-->}_{(a)} "O";"B" };
{\ar@/^12pt/^{(3)} "O";"C" };
{\ar@2{=}^{\stackrel{\text{\tiny families}}{\text{\tiny index}}}_{(4)} "C";"D" };
{\ar@{-->}_{\text{localization}}^{(c')} "L";"C" };
{\ar@{-->}_{\text{Fubini}}^{(b)} "B";"A" };
{\ar@{-->}_{\text{shriek}}^{(c)} "A";"L" };
\endxy
\end{equation}
The dashed arrows refer to path integral operations while the solid
arrows indicate procedures in the hamiltonian or operator
description. To prove the families index theorem one usually performs
the following steps in a straightforward manner.
\begin{description}
    \item[(1)]  The Atiyah-Singer index theorem tells us that on $Z$
    $\ind\slashed{D}_{Z} = \int_{Z}\Aroof(TZ)$.
  \item[(2)] Any connection can be used to compute the integral but
    the computation simplifies when the family is described by a
    riemannian submersion.  Using $T_{z}Z = H_{z} \oplus V_{z}$, the
    reduced structure group and the special
    connection~\cite{ONeill-Submersion:1966}, we see that
    $\Aroof(T_{z}Z) = \Aroof(H_{z}) \wedge \Aroof(V_{z})$.  Moreover
    the properties of the submersion connection allows us to identify
    $ \Aroof(H_{z})$ with $\Aroof(T_xX)$ which is intrinsically
    defined on the base $X$. This implies by Fubini's theorem that
    \begin{equation}
      \label{eq:fubini-rhs}
	\ind(\slashed{D}_{Z}) = \int_{X} \Aroof(TX)
	\wedge \int_{Y_{x}} \Aroof\left(T_{(x,\cdot)}Y\right)\,.
    \end{equation}
  \item[(3)] There is a second way of determining
    $\ind\slashed{D}_{Z}$.  Using our riemannian submersion geometry
    on $Z$ we can show that solving for the zero modes of
    $\slashed{D}_{Z}$ is the same as the following procedure. First
    determine the zero modes of $\slashed{D}_{Y}$ at fixed $x\in X$.
    Next solve a modified Dirac equation on $X$. The corresponding
    Dirac operator is coupled to a vector bundle with a connection
    constructed with the zero modes of $\slashed{D}_{Y}$.  The
    Atiyah-Singer index theorem tells you that the index of this
    operator is given by the integral in the lower left hand corner.
  \item[(4)] The cohomological formula for the families index theorem
    of Atiyah and Singer follows from the identification of these two
    ways of computing $\ind\slashed{D}_{Z}$ .
\end{description}
This procedure is, \emph{mutatis mutandis}, the one used by
Bismut~\cite{Bismut:1985p1131} and Bismut and
Freed~\cite{Bismut:1986p1613}.  However, as we do not know how to
generalize step (3) in the Dirac-Ramond case, we adopt a different
strategy that bypasses step (3).  This new proof relies on the path
integral computations associated with the dashed arrows of the
diagram.  Every step now has a natural extension to the loop space,
\emph{i.e.}, string theory.

We begin with $\ind\slashed{D}_{Z}$ at the center of the diagram. The
index is as usual given by the supertrace $\Tr (-1)^F$ in an
appropriate field theory where $(-1)^{F}$ is  fermionic parity.
\begin{description}
  \item[(a)] $\Tr (-1)^F=\ind\slashed{D}_{Z}$ is given by a
  supersymmetric path integral over the loop space $\loopx Z$ with
  appropriate boundary conditions
  \cite{AlvarezGaume:1983p1985,Friedan:1984p1942,Witten:AS}.  This is
  a direct result of the equivalence between the hamiltonian and the
  lagrangian formulations of quantum mechanics.
 
\item[(a')] The path integral calculation localizes on the
  ``constant loops'' in $\loopx Z$, \emph{i.e.}, the manifold
  $Z$. Steps (a) and (a') are equivalent to step (1) above and
  constitute the standard path integral derivation of the
  Atiyah-Singer index formula and we can then continue 
  using Fubini's theorem (arrow (2)).

\item[(b)] As a result of the submersion geometry, see
  \eqref{eq:metric-Z}, the lagrangian $L_{Z}$ splits naturally into
  two pieces and one of them depends only on the base $X$.  The Fubini
  theorem can be used to factorize the path integral into two factors.
  Here ``$\loopx Y$'' is the inverse image of the projection $\pi:Z\to
  X$ of a loop in $X$.
    
\item[(c)] This is the main and the only delicate step needed to
  derive the cohomological formula for the 
  families index in the Dirac-Ramond case and constitutes
  one of the main results of this paper.  The path integral over the
  fiber is the reflection in the path integral of the shriek map in
  K-Theory \cite{Atiyah-Singer-IV:1971}.  To compute the $Y$ path
  integral we notice that it satisfies a time dependent
  super-Schr\"{o}dinger equation. We then prove a new theorem in
  supersymmetric quantum mechanics that shows that this path integral
  over the fiber is exactly given by the supersymmetric parallel
  transport term in between the parentheses.  The result is the
  standard path integral expression of the index for the Dirac
  operator coupled to a bundle with connection $A$ and curvature 
  $F$~\cite{Friedan:1984p1942}.
  By construction this bundle is the index bundle.  This is the
  argument that allows us to bypass step (3).

\item[(c')] The path integral calculation localizes on ``constant
  loops'' in $X$, \emph{i.e.}, the manifold $X$, and the result of the
  computation is the A-roof genus times the Chern character of the
  $\Ind \slashed{D}_{Y}$. This step is standard and well known.

\item[(4)] \emph{cf. ibidem}
\end{description}

We can now justify the approximations we will be performing.  Our
starting point for the derivation of the families index theorem is the
computation of the index of the Dirac operator on the riemannian
submersion $Z$.  The index is an integer and therefore cannot change
as we deform the space $Z$.  If we fix a loop in the base $X$, the $Y$
path integral (in step (c)) is $\Tr (-1)^{F} U_{Y}(T,0)$ where
$U_{Y}(t,\tau)$ is the time evolution operator on $Y$ and $F$ is 
fermion number.  The time
development is obtained from a study of the super-Schr\"{o}dinger
equation and is summarized in \eqref{eq:super-holonomy}.  We will show
that in the computation of the supertrace there is an exact
cancellation and the only contribution comes from zero modes.  Since
the $\ind Z$ is an integer we can go to a parameter regime where the
adiabatic approximation is valid and in this way we clarify the
contribution of the zero modes.  In the adiabatic approximation, the
travel time $T$ around a loop is taken to be very large, and the
riemannian submersion metric is blown up in such a way that the
evolution is very slow as one goes from $t=0$ to $t=T$.  The
contribution from the zero modes in the adiabatic approximation is
given by super-parallel transport in the index bundle around the loop
in $X$.  Our remarks imply that this is an exact result.

\subsection*{Details of step (c).}
\label{sec:step-c}

The proof of step~(c) in diagram~\eqref{eq:diag} requires a discussion
of the definition of the super-heat kernel in supersymmetric quantum
mechanics.  The standard framework is the following.  On a $(1|1)$
super-manifold with coordinates $(t,\tau)$, where $\tau$ is a
Grassmann variable, the supersymmetry transformation acts as $t \to t
+ i\epsilon \tau$ and $\tau \to \tau + \epsilon$.  The generator of
supersymmetry is $Q = \partial_{\tau} + i\tau \partial_{t}$ with
$Q^{2} = i\partial_{t}$.  The superderivative is $D = \partial_{\tau}
-i\tau\partial_{t}$ with $D^{2}=-i\partial_{t}$ and anti-commutator
$\{D,Q\}=0$.  After quantization $Q$ becomes an operator on the
Hilbert space and we will interpret $Q$ this way from now on.  The
fundamental solution of the super-Schr\"{o}dinger equation
\begin{equation}
  \label{eq:Super-Sch}
  D\Phi(t,\tau) = Q\Phi(t,\tau)
\end{equation}
is the super-heat kernel.  This equation was introduced in
\cite{Friedan:1984p1942} to study the index of the Dirac operator in
an intrinsically supersymmetric covariant manner.  We are interested
in a generalization of the above that is analogous to going from a
time independent hamiltonian to a time dependent one.  We are
interested in solving \eqref{eq:Super-Sch} where we have $(t,\tau)$
dependence, \emph{i.e.}, $Q(t,\tau)=Q_{0}(t)+\tau Q_{1}(t)$.  Note
that in this case $\{D,Q\} \neq 0$.  The fundamental solution to the
super-Schr\"{o}dinger equation with initial value $U_{Y}(0,0)=I$ is
the super-heat kernel $U_{Y}(t,\tau)$.  The equivalence of the
operator and the path integral formulations of quantum mechanics tells
us that with supersymmetric boundary conditions we have
\begin{equation}
    \Tr (-1)^{F} U_{Y}(T,0) = \int_{\pi^{-1}(\gamma)} e^{L_{Y}}\,.
    \label{eq:path-int-Y}
\end{equation}
Here $\pi^{-1}(\gamma)$ is the inverse image under $\pi:Z\to X$ of a 
superloop $\gamma$ on $X$. One of our key results  is that the 
left hand side of the equation above is exactly given by the
super-holonomy on the index bundle around the superloop 
$\gamma$, the generalization of the heat kernel expression for 
the index to the families case.
We can expand the wavefunction as
\begin{equation}
\label{eq:Phi_expanded}
    \Phi(t,\tau) = \sum_{n} \phi_{n}(t,\tau) b_{n}(t,\tau)\, ,
\end{equation}
where $\{\phi_{n}\}$ will be taken to be a complete orthonormal basis
with $Q(t,\tau) \phi_{n}(t,\tau) = \lambda_{n}(t) \phi_{n}(t,\tau)$.
The eigenfunctions of $Q$ can be constructed if we know the
eigenfunctions of $Q_{0}$.  Let $\phi_{0}$ be an eigenfunction of
$Q_{0}$, $Q_{0}\phi_{0} = \lambda\phi_{0}$ then $\phi=
\phi_{0}-(Q_{0}-\lambda)^{-1}\tau Q_{1}\phi_{0}$ is an eigenfunction
of $Q$ with eigenvalue $\lambda$.  The resolvent is defined to vanish
on $\ker(Q_{0}-\lambda)$.
Inserting \eqref{eq:Phi_expanded}
into \eqref{eq:Super-Sch}  and 
taking the inner product with $\phi_{m}$ gives  the exact equation
\begin{equation}
    Db_{m}  + \sum_{n} \bigl( \phi_{m},D\phi_{n} \bigr) b_{n} = \lambda_{m} 
    b_{m}\,.
    \label{eq:sup-ad-1}
\end{equation}
Notice that the super-Schr\"{o}dinger equation \eqref{eq:Super-Sch} is
very similar to the equation which defines super-parallel transport.
Assume we have a $(1|1)$ superparticle moving on a manifold $M$ where
there is a non-abelian connection $A$.  The motion of the particle is
described by the superfield $X^{\mu}(t,\tau) = x^{\mu}(t) + i\tau
\xi^{\mu}(t)$. The manifold $X$ is not to be
confused with the superfield $X(t,\tau)$ that describes a superloop
$\gamma$ on $X$. The super-parallel transport equation,
\begin{equation}
    D\Phi(t,\tau) + A_{\mu}(X)DX^{\mu}\;\Phi(t,\tau)=0m
    \label{eq:super-parallel}
\end{equation}
is a multi-component super-Schr\"{o}dinger equation with $Q(t,\tau) 
= -A_{\mu}(X)DX^{\mu}$. If we 
split the super-parallel transport into bosonic and fermionic 
components, with  $\Phi(t,\tau) = \Phi_{b}(t) + i\tau \Phi_{f}(t)$, we obtain
\begin{align}
    \dot{\Phi}_{b} + \left( A_{\mu}\dot{x}^{\mu} - \tfrac{i}{2} 
    F_{\mu\nu}\xi^{\mu}\xi^{\nu} \right) \Phi_{b} &=0\,,
    \label{eq:spt-b}  \\
    \Phi_{f} + A_{\mu}\xi^{\mu}\Phi_{b} & =0\,,
    \label{eq:spt-f}
\end{align}
where $F_{\mu\nu}=\partial_{\mu}A_{\nu}-\partial_{\nu}A_{\mu} +
[A_{\mu}, A_{\nu}]$.  Super-parallel transport is ordinary parallel
transport with an extra ``rotation'' given by a Pauli
$\vec{\sigma}\cdot \vec{B}$ type coupling.  Note that $\Phi(t,\tau) =
\left(1 - i\tau A_{\mu}(x)\xi^{\mu} \right) \Phi_{b}(t) $ and thus
$\Phi(t,\tau=0) = \Phi_{b}(t)$.  We  use this later when we apply
the same methodology to \eqref{eq:sup-ad-1}.

The key point is that we only have to compute the supertrace of
$U_{Y}(t,0))$ and not the full operator.  The term
$(\phi_{m},D\phi_{n})$ in \eqref{eq:sup-ad-1} gives a supersymmetric
Berry-Simon connection~\cite{Simon:1983p3235}. In the type of systems
we are studying, $Q$ depends on $(t,\tau)$ implicitly through a
bosonic superfield $X(t,\tau)$, with $\tau X(t,\tau) = X(t,\tau)\tau$
which is apparent from the form of the lagrangian $L_{Z}$.  We have
$\phi_{n} \bigl( X(t,\tau) \bigr)$ and $D\phi_{n} = (\partial
\phi_{n}/\partial x^{\mu})(X) \; DX^{\mu}\,.$ With this in mind we
conclude that $(\phi_{m},D\phi_{n}) = A^{mn}_{\mu}(X)\; DX^{\mu}$
where $A^{mn}_{\mu}(X) = \left( \phi_{m}(X), (\partial
\phi_{n}/\partial x^{\mu})(X) \right)$.  Thus equation
\eqref{eq:sup-ad-1} may be written as
\begin{equation}
    Db_{m} + \sum_{n} A^{mn}_{\mu}(X)\,DX^{\mu}\, b_{n} = \lambda_{m}
    b_{m}\,.
    \label{eq:sup-ad-4}
\end{equation}
Let $\Lambda$ be the matrix with the $\lambda_{n}$ on the
diagonal.  Using \eqref{eq:spt-b} and \eqref{eq:spt-f} we  find
$b(t,\tau) =\left[ 1- \tau \bigl(A_{\nu}(x(t))\xi^{\nu}(t) +
i\Lambda(t) \bigr) \right] b^{b}(t)$ where the bosonic component
$b^{b}$ of $b(t,\tau) = b^{b}(t) + i\tau b^{f}(t)$ satisfies
\begin{align}
    \dot{b}^{b} + \left\{ i \Lambda^{2} 
    + \left( A_{\mu}\dot{x}^{\mu} - \tfrac{i}{2}\, 
    F_{\mu\nu}\xi^{\mu}\xi^{\nu} \right) 
    + \left[ A_{\mu}, \Lambda \right] \xi^{\mu} \right)\} b^{b} =0\,,
    \label{eq:bb-1}
  \end{align}
and contains all the information necessary to determine $U_{Y}(T,0)$.
We choose positive integers $n$ to label the orthonormal eigenvectors
$\phi_{0n}=\varphi_{n}$ of $Q_{0}$ with eigenvalue $\lambda_{n}>0$.
The eigenvector $\varphi_{-n} = (-1)^{F}\varphi_{n}$ has eigenvalue
$-\lambda_{n}<0$.  The zero modes of $Q_{0}$ are indexed by $z$ and
denoted by $\varphi_{z}$.  A detailed analysis of \eqref{eq:bb-1}
shows that $b^{b}_{n}(t)$ and $b^{b}_{-n}(t)$ satisfy the same
differential equation when the action of $(-1)^{F}$ is taken into
account.  Therefore we have an exact cancellation in the supertrace
$\Tr (-1)^{F}U_{Y}(T,0)$ from the states orthogonal to $\ker Q_{0}$.
Since the evolution is given by a first order differential equation
this argument can be applied at each instant of time and can be
adapted to the case when the kernel jumps.  We still have to compute
the contribution of the zero modes to the supertrace.  This can be
done exactly by the adiabatic approximation that we review shortly.

Applying these ideas to the zero modes we will obtain a refinement to
the adiabatic theorem. Within the adiabatic approximation,
the amplitudes $b_{z}$ for the zero modes satisfy the equation
\begin{equation}
    Db_{z}  + \sum_{z'} A^{z,z'}_{\mu}(X) (DX^{\mu}) b_{z'} = 0
    \label{eq:sup-ad-3}
\end{equation}
that is the super-parallel transport
equation~\eqref{eq:super-parallel}.  This gives the exact result
\begin{equation}
    \Tr (-1)^{F} U_{Y}(T,0) = \Tr (-1)^{F} \Phi_{b}(T)\,,
    \label{eq:super-holonomy}
\end{equation}
where $\Phi_{b}$ is the superholonomy given by the super-parallel
transport equation~\eqref{eq:spt-b}.  This is the refinement of
Berry's phase to supersymmetric quantum mechanics.  This superparallel
transport is now an additional term that has to be added to
supersymmetric lagrangian on $X$.  It reflects a coupling of the
superparticle on $X$ to a gauge field on the index bundle.  Next, we
can perform the $Y$ path integral in step (c) and as usual the $X$
path integrals in step (c') and express $\ind \Dirac_{Z}$ as
\begin{equation}
  \label{eq:shriek+loc}
\int_{\loopx X} e^{L_{X}} \left(e^{\int A + \int
      F}\right) =  \int_{X} \Aroof(TX) \ch\left(\Ind
    \Dirac_{Y}\right)\,
\end{equation}
If we compare  \eqref{eq:shriek+loc} and
\eqref{eq:fubini-rhs}, we get
\begin{equation*}
    \int_{X} \Aroof(TX) \left( \chern(\IND(\slashed{D}^{Y})) -
    \int_{Y_{x}} \Aroof\left(T_{(x,\cdot)}Y\right) \right)=0.
\end{equation*}
Thus we have that as cohomology classes on $X$
\begin{equation}
    \left[\chern \left(\IND(\slashed{D}^{Y})\right)\right] = \left[
    \int_{Y_{x}} \Aroof\left(T_{(x,\cdot)}Y\right)\right],
    \label{eq:fam-ind-1}
\end{equation}
where $[\;\cdot\;]$ denotes the cohomology equivalence class.  This is
the cohomological families index formula for the Dirac operator which
ends our discussion of the the Dirac operator case.

\subsection*{Adiabatic approximation with symmetries.}
\label{sec:adiabatic}

We now turn our attention to the Dirac-Ramond operator.  The new
feature is that $P(t)$, the spatial translation operator along the
loop, commutes with the hamiltonian $H(t)$.  Our analysis requires the
use of the adiabatic approximation in the presence of symmetries, a
topic that is not addressed in textbooks.  To simplify the discussion
supersymmetry will be ignored. In the adiabatic approximation we
scale the time so that the Schr\"{o}dinger equation becomes
$i\,\partial \psi/\partial t = T H(t) \psi(t)\,.  $ Let $\{\varphi_{ r
}(t)\}$ be an orthonormal basis of eigenvectors of $H(t)$ with
eigenvalue $E_r(t)$.  The orthonormal basis expansion for
wavefunctions will be written as
\begin{equation*}
    \psi(t) = \sum_{ r } e^{-i T \int_{0}^{t} E_{ r }(t') 
    \;dt'}\; a_{ r }(t) \varphi_{ r }(t)\,.
    \label{eq:S-soln}
\end{equation*}
Inserting this into the Schr\"{o}dinger equation
and taking the inner product with $\varphi_{ s }$ gives the exact 
equation
\begin{equation}
    \dot{a}_{ s }(t) + \sum_{ r } e^{iT 
    \int_{0}^{t}(E_{ s }(t') -E_{ r }(t'))\;dt'} \; 
    \braket{\varphi_{ s }}{\dot{\varphi}_{ r }} a_{ r }(t)
    =0\,.
    \label{eq:exact-soln}
\end{equation}
As $T \to\infty$ an oscillating term contributes very
little~\cite{Messiah:1964p1854} and the states $\varphi_{ s }$ with
$E_{ s }(t) = E_{ r }(t)$ for all $t>0$ are the only ones needed.  In
general there will only be one such state except in cases in which a
symmetry enforces a multiplicity.  In these cases one obtains the
adiabatic approximation result $\dot{a}_{ s }(t) + \sum_{r, E_{ r
}=E_{ s }} \braket{\varphi_{ s }}{\dot{\varphi}_{ r }} a_{ r }(t)
\approx 0$ within a degenerate energy level.  On the eigenspace with
eigenvalue $E_{ s }$ we get a connection along the family of
hamiltonians given by $A_{ s r }(t) = \braket{\varphi_{ s
}}{\dot{\varphi}_{ r }}$.  The holonomy of this connection is the
non-abelian Berry's phase \cite{Wilczek:1984p2825}.  This connection
is not unitary because the volume element of the fiber $Y_{x}$ can
change as $x$ varies \cite{Bismut:1986p2775}.  The hermitian piece of
the connection is associated with the varying volume element and takes
the form: $\int_{Y_{x}}dy \; \sqrt{g_{Y}} \Tr(g_{Y}^{-1}\dot{g}_{Y})
\varphi_{s}^{*}\varphi_{r}$.  Standard perturbation theory
computations show that the connection has an ``irreducible'' part and
a ``perturbative'' part.  The topological information is contained in
the ``irreducible'' part while the ``perturbative'' part, a
differential form of type $ad$, corresponds to a translation in the
affine space of connections.  The perturbative parts can be ignored
when discussing topological invariants.  Thus we can ignore the fiber
volume related part of the connection.  The parts of the connection
associated to transitions between different eigenvalues of $H(t)$ are
purely ``perturbative'' because
$\braket{\varphi_{s}}{\dot{\varphi}_{r}} =
\bra{\varphi_{s}}{\dot{H}}\ket{\varphi_{r}}/(E_{r}-E_{s})$.  Moreover
if we write an orthogonal direct sum for the Hilbert space
$\mathcal{H} = \bigoplus_{E(t)} \mathcal{H}_{E(t)}$, in terms of the
eigenspaces of $H(t)$ then the ``irreducible'' part of the connection
is a direct sum $A = \bigoplus_{E(t)} A_{E(t)}$ where each piece
$A_{E(t)}$ may be taken to be a unitary $\U(\dim \mathcal{H}_{E(t)})$
connection.

We can extend this analysis to a theory with additional symmetries
because we have two commuting symmetries in the study of the
Dirac-Ramond operator.  Here we only consider the case of a maximally
commuting algebra of self-adjoint operators $\mathcal{C}(t)$.  Its
basis will be written as $\{ H(t)=H_{0}(t), H_{1}(t), \ldots,
H_{l}(t)\}$.  We assume that the spectrum of any operator in
$\mathcal{C}(t)$ is discrete.  Since $\mathcal{C}(t)$ is abelian, its
irreducible representations are one dimensional.  We can find
simultaneously eigenvectors of all the operators in $\mathcal{C}(t)$.
A state $\psi$ has weight $\lambda(t)$ if $H_{i}(t) \psi =
\lambda_{i}(t) \psi$ where $\lambda(t) = (E(t), \lambda_{1}(t),\ldots,
\lambda_{l}(t))$.  We will express the Hilbert space as an orthogonal
direct sum $\mathcal{H} = \bigoplus_{\lambda(t)}
\mathcal{H}_{\lambda(t)}$.  As $t$ varies, the subspaces
$\mathcal{H}_{\lambda(t)}$ also vary.  Assume that
$\varphi_{\lambda}(t)$ and $\varphi_{\mu}(t)$ are normalized
eigenvectors with respective weights $\lambda(t)$ and $\mu(t)$, and
$\lambda(t) \neq \mu(t)$.  It follows that there exists $j$ such that
$\lambda_{j}(t) \neq \mu_{j}(t)$.  Applying the same analysis as above
to the operator $H_{i}(t)$, one concludes on the one hand that the
connection between the subspaces $\mathcal{H}_{\lambda(t)}$ and
$\mathcal{H}_{\mu(t)}$ is perturbative and on the other hand that the
``irreducible'' part of the connection is a direct sum $A =
\bigoplus_{\lambda(t)} A_{\lambda(t)}$ where each piece
$A_{\lambda(t)}$ is a $\U(\dim \mathcal{H}_{\lambda(t)})$ connection
on $\mathcal{H}_{\lambda(t)}$.

For simplicity consider a situation with only two commuting operators
$H(t)$ and $P(t)$ and where the spectrum of $P(t)$ takes integer
values.  For the adiabatic time evolution by the operator $H(t) +
\theta P(t)/T$, the contribution from a subspace of the Hilbert space
with energy $E$ and with $P$ eigenvalue $n$ is schematically given by
$e^{-i\theta n} \Tr e^{-\int A}$, where $A$ is the Berry-Simon
connection on that subspace.

In the Dirac-Ramond model we have a $(0,1)$ supersymmetry.  This means
that there are operators $L_{0}$, $\bar{L}_{0}$ that commute with each
other and that the Dirac-Ramond operator satisfies $\bar{G}_{0}^{2} =
\bar{L}_{0} - \bar{c}/24$.  Its index is given by the supertrace
$\Tr(-1)^{F} q^{L_{0}-c/24} \bar{q}^{\bar{G}_{0}^{2}}$.  To relate
this to the previous discussion we note that the time development
operator is $H = (L_{0}-c/24) + (\bar{L}_{0} - \bar{c}/24)$ and the
spatial translation operator is $P = (L_{0}-c/24) - (\bar{L}_{0} -
\bar{c}/24)$. Applying the previous analysis we conclude that the
result of step (c) in the Dirac-Ramond scenario is an expression of
the form $\sum_{n=0}^{\infty} q^{n-c/24} \;\Phi_{n}$ where $\Phi_{n}$
is the superholonomy of the index bundle at ``$n$-th level
eigenspace'' of $L_{0}-c/24$.  Step (c') gives localization and will
lead to formula \eqref{eq:lhs}.

\subsection*{The families index for Dirac-Ramond.}
The existence of a Dirac-Ramond operator on $Z$ requires $\pont(Z)=0$
and $\dim Z = n =4\nred$ while the submersion structure implies by
restriction that $\pont{Y}=0$. The family defined by $Z$ has base $X$
with $\dim X =p$ and fiber isomorphic to $Y$ with $\dim Y=m$.  There
will be two distinct cases to consider.  The first case is $p=4\pred$
and $m=4\mred$ where $\pred \ge 0$ and $\mred \ge 1$. The second case
is $p=4\pred+2$ and $m=4\mred-2$ where $\pred \ge 0$ and $\mred \ge
1$.  Note that in both cases $n=4\nred=4(\pred+\mred)$.

In the Dirac-Ramond case with (0,1) supersymmetry, the
right hand side \eqref{eq:fam-ind-1} becomes
\begin{equation}
    \int_{Y_{x}} \sgenus(\Omega_{Y},\tau) = \frac{1}{\eta(\tau)^{m}}
    \int_{Y_{x}} \aroof(\Omega_{Y},\tau)\,,
    \label{eq:rhs}
\end{equation}
where $Y_{x}$ is the fiber of $Z\to X$ over $x\in X$ and $\Omega_{Y}$
is the curvature $2$-form using the submersion connection of the
vertical tangent bundle. The notations we use for the string genus
$\sgenus$ and for $\aroof$ are summarized in the appendix. From the
path integral point of view, the Dirac and the Dirac-Ramond case
differ by the essential presence of a full right Virasoro algebra
commuting with supersymmetry.  In particular since $[L_{0},\susy]=0$,
the index bundle for the family defined by $Z\to X$ has an orthogonal
direct sum decomposition into representations of $L_{0}$ given by
$\IND(Z\to X) = \bigoplus_{n=0}^{\infty} \IND_{n}(Z\to X)$, where $n$
denotes the grading with respect to $L_{0}$.  Our adiabatic invariance
arguments in the presence of an abelian symmetry show that this leads
to a  connection $A = \bigoplus_{n=0}^{\infty} A^{(n)}$ where
$A^{(n)}$ acts only on $\IND_{n}(Z\to X)$.  There are no off diagonal
pieces that connect subspaces labeled by different values of $n$.  We
denote the curvature of $A^{(n)}$ by $F^{(n)}$.  The left hand side of
the formula for the family's index theorem is given by what we call
the graded string Chern character
\begin{equation}
    \echern(\tau,F)=
    \sum_{n=0}^{\infty} q^{n-m/24}\; 
    \chernchar \left(iF^{(n)}/2\pi \right),
    \label{eq:lhs}
\end{equation}
where $\chernchar$ is the Chern character. 

Given the foregoing, we are in a position to write down our main
result:
\begin{equation}
    \echern(\tau,F) = \left[\int_{Y_{x}} \sgenus(\Omega_{Y},\tau) \right]\,.
    \label{eq:cargese-0}
\end{equation}
This formula, the equivalent of \eqref{eq:fam-ind-1} for loop space,
gives a families index theorem for the Dirac-Ramond operator. Its
right hand side can be rewritten in a more explicit way.  Let $\sgenus
= \sum_{k=0}^{\infty} \sgenus_{4k}(\Omega_{Y},\tau)$, and $\aroof=
\sum_{k=0}^{\infty} \aroof_{4k}(\Omega_{Y},\tau)$ where $\sgenus_{4k}$
and $\aroof_{4k}$ are $4k$-forms on $Z$.  Note that
$\aroof_{4k}(\Omega_{Y},\tau) = \eta(\tau)^{m}\,
\sgenus_{4k}(\Omega_{Y},\tau)$.  It is convenient to define
$\aroof_{l} = 0$ if $l \not\equiv 0 \bmod 4$ and likewise for
$\sgenus_{l}$.  Integration along the fiber reduces the $4k$-form to a
$4k-m$ form on the base $X$
\begin{equation}
\begin{split}
    \int_{Y_{x}} \aroof_{4k}(\Omega_{Y},\tau) = \alpha_{4k-m}(x,\tau)\\
    \int_{Y_{x}} \sgenus_{4k}(\Omega_{Y},\tau) = \sigma_{4k-m}(x,\tau).
    \label{eq:fiber-integration}
  \end{split}
\end{equation}
Note that $\alpha_{l}(x,\tau) = \eta(\tau)^{m}\, \sigma_{l}(x,\tau)$,
and if $l <0$ then $\alpha_{l}=0$, $\sigma_{l}=0$.  It then follows
from the modular properties of $\aroof$ that $\alpha_{4k-m}$
transforms as a modular form of weight $2k$  while 
\begin{align}
    \sigma_{4k-m}(\tau+1) &  = e^{-2\pi i m/24}\, \sigma_{4k-m}(\tau)\,,
    \label{eq:sigma-T}  \\
    \sigma_{4k-m}(-1/\tau) & = e^{2\pi i m/8}\,\tau^{(4k-m)/2}\,
    \sigma_{4k-m}(\tau)\,.
    \label{eq:sigma-S}
\end{align}
More precisely if we take $4k-m$ vectors $X_{1},\ldots,X_{4k-m} \in T_{x}X$ then
$\alpha_{4k-m}(x,\tau)(X_{1},\ldots,X_{4k-m})$ will be a modular form
in $\tau$ of weight $2k$.
Observe that $0 \le 4k-m \le p=\dim X$ and
therefore $\dim Y = m \le 4k \le p+m = \dim Z$.  The range for the
weight of the modular form is $\tfrac{1}{2} \dim Y \le 2k \le
\tfrac{1}{2} \dim Z$.  We can make the weight of $\alpha_{4k-m}$ as
large as possible by making the parameter space $X$ have high
dimensionality.
All this may be summarized in the formulae
\begin{align}
    \echern(\tau+1,F) &= e^{-2\pi i m/24}\, \echern(\tau,F)\,,
    \label{eq:ech-T}\\
    \echern(-1/\tau,F/\tau) &= e^{2\pi i m/8}\, \echern(\tau,F)\,.
    \label{ech-S}
\end{align}
This extends the modular properties of the index
\eqref{eq:sgenus-int-mod} to the full index bundle.  If we write
$\echern(\tau,F) = \alpha(\tau,F)/ \eta(\tau)^{m}$ then
\begin{align}
    \alpha(\tau+1,F) &= \alpha(\tau,F)\,, 
    \label{eq:alpha-T}\\
    \alpha(-1/\tau,F/\tau) &= \tau^{m/2}\alpha(\tau,F)\,.
    \label{eq:alpha-S}
\end{align}
If $y^{(n)}_{j}$ are the formal eigenvalues of
$iF^{(n)}/2\pi$ then $\chernchar(iF^{(n)}/2\pi) =
\sum_{j}e^{y^{(n)}_{j}}$.  The Chern character is formally invariant
under the transformation $y^{(n)}_{j} \to y^{(n)}_{j} + 2\pi i
m^{(n)}_{j}$ where $m^{(n)}_{j} \in \bbZ$ and therefore $\alpha(\tau,
y^{(n)}_{j})$ should be periodic 
\begin{equation}
    \alpha(\tau, y^{(n)}_{j})=
    \alpha(\tau, y^{(n)}_{j}+ 2\pi i m^{(n)}_{j})
    \label{eq:alpha-real}
\end{equation}
Combining this with \eqref{eq:alpha-T} and \eqref{eq:alpha-S} we see 
that there is an additional formal periodicity
\begin{equation}
    \alpha(\tau, y^{(n)}_{j})=
    \alpha(\tau, y^{(n)}_{j}+ 2\pi i l^{(n)}_{j}\tau)
    \text{ where } l^{(n)}_{j}\in\bbZ.
    \label{eq:alpha-tau}
\end{equation}
These formal transformation properties of the $\alpha$ remind one of
the transformation rules of Jacobi forms.  However this formal
transformation cannot be a true invariance.  One way to see this is to
fix a positive integer $r$ and consider $y^{(n)}_{j} \to y^{(n)}_{j} -
2\pi i r \delta_{nr}$.  This transformation completely eliminates the
$q^{r-m/24}$ term from \eqref{eq:lhs}.  We conclude that the
$y^{(n)}_{j}$ are not all independent as can be seen clearly in the
particular cases studied  in \eqref{eq:app-1} and
\eqref{eq:app-2} below.  For example \eqref{eq:app-1} implies that all
the first Chern classes are related and determined by the single Chern
class allowed in the right hand side.

The modular properties of $\sigma_k$ severely constrains the non
trivial cohomology classes of the index bundle of the Dirac-Ramond
operator.  We rewrite \eqref{eq:cargese-0} as
\begin{equation}
    \sum_{n=0}^{\infty} q^{n-m/24}\; 
    \chernchar \left(iF^{(n)}/2\pi \right) = \sum_{k=0}^{\infty} 
    \sigma_{4k-m}(\tau)\,
    \label{eq:cargese}
\end{equation}
and the classes are  described by a universal
formula. Let $\dim Y = m= 4\mred -2\varepsilon$ where
$\varepsilon=0,1$ then the $4j+2\varepsilon$ cohomology class of the
index bundle is given by
\begin{align}
  \frac{1}{(2j+\varepsilon)!} \sum_{n=0}^{\infty} q^{n-m/24}\;
  \Tr\left(\frac{iF^{(n)}}{2\pi}\right)^{2j+\varepsilon}  =
  \sigma_{4j+2\varepsilon}(\tau) 
\end{align}
where 
\begin{equation}
  \sigma_{4j+2\varepsilon}(\tau) = 
  \frac{\alpha_{4j+2\varepsilon}(\tau)}{\eta(\tau)^{m}}
  =\int_{Y_{x}}
  \sgenus_{4j+4\mred}(\Omega_{Y},\tau) 
  \label{eq:universal-coho}
\end{equation}
Note that $\alpha_{4j+2\varepsilon}$ is of modular weight $2j+2\mred$.
An important reminder is that with our conventions $\mred \ge 1$ and
therefore there is a bound on the modular weight of
$\alpha_{4j+2\varepsilon}$ given by $2j + 2\mred \ge 2j +2$.

\subsection*{Some applications.}
\label{sec:some-results}
Two different approaches come to mind when trying to exploit the
cohomological formula \eqref{eq:universal-coho}. On the one hand we
can use what we know about $\dim M_{k}$, the dimensionality of the 
space of modular forms of weight $k$, and in particular look at
spaces of modular form of low dimensionality. This restricts
$\alpha_{4j+2\epsilon}$ and can place strong constraints on the
cohomological classes. Alternatively we can study cohomology of a
specified degree.

As a first example we look at the case where
$\alpha_{4j+2\varepsilon}$ has modular weight $4$ . There are two
possibilities for $(j,\mred)$ given by $j=0$, $\mred=2$; and $j=1$,
$\mred=1$.  Because $\dim M_{4}=1$ we have that
$\alpha_{4j+2\varepsilon}(x,\tau) = E_{4}(\tau)
\tilde{\alpha}_{4j+2\varepsilon}(x)$ where $\tilde{\alpha}$ a closed
$(4j+2\varepsilon)$-form on $X$ that is independent of $\tau$ and 
$E_{4}$ is the Eisenstein series.  When
$\epsilon=1$, $\dim Y =4\mred -2$.  If $j=0$, $\mred=2$ then we are
looking at the second cohomology of the index bundle for the
Dirac-Ramond operator on a manifold of dimension $\dim Y = 4\mred-2=6$
and we can take $X$ to be a two dimensional manifold.  We see that
\begin{equation}
   \sum_{n=0}^{\infty} q^{n-1/4} \Tr\left(
    \frac{iF^{(n)}}{2\pi}\right) =
    \frac{E_{4}(\tau)}{\eta(\tau)^{6}}\; \Tr\left(
    \frac{iF^{(0)}}{2\pi}\right)\,.
    \label{eq:app-1}
\end{equation}
The first Chern class of the determinant line bundle of $\IND_{n}$ is
proportional to the first Chern class of the determinant line bundle
of the Dirac operator.  If you eliminate the ``anomaly'' associated
with the Dirac operator on $Y$ then you eliminate the anomaly for the
Dirac operator coupled to appropriate powers of $TY$.  The other case
has $j=1$ with $\dim Y=4\mred-2=2$ where we reach the conclusion that
the $6$-cohomology of the index bundle (need $\dim X \ge 6$) is given
by
\begin{equation}
   \sum_{n=0}^{\infty} q^{n-1/12} \Tr\left( 
   \frac{iF^{(n)}}{2\pi}\right)^{3} = 
   \frac{E_{4}(\tau)}{\eta(\tau)^{2}}\; \Tr\left( 
   \frac{iF^{(0)}}{2\pi}\right)^{3}\,.
   \label{eq:app-2}
\end{equation}

An interesting example of the second type from a physics point of view
is given by two cohomology.  This case corresponds to $j=0$,
$\varepsilon=1$ and gives the first Chern class of the determinant
line bundles associated with the various index bundles $\IND_{n}$. The
manifold $Y$ has dimension $m=4\mred-2$.  We can take $X$ to be a
$2$-manifold.  The $2$-form $\alpha_{2}$ has weight $2\mred = \dim
Y/2+1$ so that if $\dim Y = 6,10,14,18,26$ (respectively
$2\mred = 4, 6, 8, 10, 14$) then $\dim M_{2\mred}=1$ and we conclude
that the first Chern classes of the index bundle are given by
\begin{equation*}
    \sum_{n=0}^{\infty} q^{n-m/24} \Tr\left(
    \frac{iF^{(n)}}{2\pi}\right) = 
    \frac{E_{2\mred}(\tau)}{\eta(\tau)^{m}}\; \Tr\left(
    \frac{iF^{(0)}}{2\pi}\right).
    \label{eq:sg-2}
\end{equation*}
In this range if the determinant line bundle of the Dirac operator has
vanishing first Chern class then so do all the determinant line
bundles for the higher operators.  In the especially interesting case
with $\dim Y=10$, \emph{i.e.}, $2\mred=6$, the result is
\begin{equation*}
    \sum_{n=0}^{\infty} q^{n-5/12} \Tr\left(
    \frac{iF^{(n)}}{2\pi}\right) = 
    \frac{E_{6}(\tau)}{\eta(\tau)^{10}}\; \Tr\left(
    \frac{iF^{(0)}}{2\pi}\right).
\end{equation*}
Hence you see that $\Tr( iF^{(1)}/2\pi) = -494 \Tr(
iF^{(0)}/2\pi)$.  $F^{(0)}$ is the curvature of the index bundle of
the Dirac operator and $F^{(1)}$ is the curvature of the index bundle
of the Dirac operator coupled to $TY$.  If you compare this to the
calculation of Alvarez-Gaum\'{e} and
Witten~\cite{AlvarezGaume:1984p1466} for the gravitational anomalies
in type IIB supergravity in a manifold with $\pont(Z)=0$ you find that
$\mathcal{A}_{3/2} = -495 \mathcal{A}_{1/2}$ where $\mathcal{A}_{1/2}$
is the anomaly contribution from the chiral spinor and
$\mathcal{A}_{3/2}$ is the contribution from the chiral gravitino.
The difference of $1$ corresponds to the longitudinal component of a
Rarita-Schwinger field $\psi_{\mu}$ that must be accounted for 
correctly to get the
physical gravitino.  Note that $\dim Z=12$ and $\pont(Z)=0$ tell us
that things can only depend on $\pontx_{3}$ so all the anomalies will
be proportional.

\subsection*{Conclusions and outlook.}

We have shown that the characteristic classes of the index bundle of
the Dirac-Ramond operator have remarkable modular properties.  The
discussion was here restricted to families described by a riemannian
submersion.  In a forthcoming longer publication we extend the
analysis to the Dirac-Ramond operator coupled to various infinite
dimensional vector bundles and show how symmetries constrain the
structure by using the representation theory of Virasoro and chiral
algebras.   An open important question is the link of our geometrical 
methods to tmf.

\subsection*{Appendix.}
\label{sec:string-mod}

The index of the Dirac-Ramond operator for $(0,1)$ supersymmetry on a
manifold $M$ with $\pont(M)=0$ is given by $\Tr_{\ker \susy} (-1)^{F}
q^{L_{0}-c/24} =\int_{M}\sgenus(M,\tau) \label{String-genus-s}$, where
$\susy$ is the generator of the right handed supersymmetry.  The
integrand is the string genus $\sgenus(M,\tau) =\aroof(M,\tau)/
\eta(\tau)^{d}$ where $d=\dim M$ and
\begin{equation}
    \aroof(M,\tau) = \prod_{j=1}^{d/2} 
    \frac{ix_{j}/2\pi}{\sigma(ix_{j}/2\pi,\tau)}\,.
    \label{eq:a-roof}
\end{equation}
Here $\eta$ is the Dedekind eta function $\eta(\tau) =
q^{1/24} \prod_{n=1}^{\infty}\left(1 - q^{n}\right)$, where $q =
e^{2\pi i \tau}$ and $\sigma$ is the Weierstrass function.
The modular transformation properties of the string genus follow from
$\aroof(x,\tau+1) = \aroof(x,\tau)$ and $\aroof(x/\tau,-1/\tau) =
\aroof(x,\tau)$ This implies for the string genus
$\sgenus(x/\tau,-1/\tau) = \sgenus(x,\tau)/(-i\tau)^{d/2}$.  Since
integration over $M$ picks out the differential form with degree $\dim
M$,
$\int_{M}\aroof(M,\tau)$ is a modular form of weight $d/2$ and 
\begin{equation}
  \begin{split}
    \label{eq:sgenus-int-mod}
    \int_{M}\sgenus(M,-1/\tau) &= e^{2\pi i d/8}\; 
    \int_{M}\sgenus(M,\tau)\;,\\
    \int_{M}\sgenus(M,\tau+1) &= e^{-2\pi i d/24}\; 
    \int_{M}\sgenus(M,\tau)\;.
  \end{split}
\end{equation}

\begin{acknowledgments} 
  We would like to thank M.~Hopkins and I.M.~Singer for extensive
  discussions and for their hospitality at Harvard and  MIT where part
  of this work was done.  We thank J.M.~Bismut for bringing
  ref. \cite{Liu:2000p3725} to our attention. We are also grateful to
  our home institutions for supporting multiple visits. The work of OA
  was supported in part by the National Science Foundation under
  Grants PHY-0244261 and PHY-0554821. The work of PW has been
  supported in part by the European Community Human Potential Program
  under contract MRTN-CT-2004-512194 and by Agence Nationale de la
  Recherche under contract ANR(CNRS-USAR) no.05-BLAN-0079-01.
\end{acknowledgments}

\relax

\end{article}
\end{document}